# Observation of a Topological Insulator Dirac Cone Reshaped by Non-magnetic Impurity Resonance

Lin Miao,[1,2][†] Yishuai Xu,[1][†] Wenhan Zhang,[3][†] Daniel Older,[1]

S. Alexander Breitweiser,[1] Erica Kotta,[1], Haowei He,[1] Takehito Suzuki,[4]

Jonathan D. Denlinger,[2] Rudro R. Biswas,[5]

Joseph Checkelsky,[4] Weida Wu,[3] L. Andrew Wray,[1,6]*

[1] Department of Physics, New York University, New York, New York 10003, USA
[2] Advanced Light Source, Lawrence Berkeley National Laboratory, Berkeley, CA 94720, USA
[3] Rutgers Department of Physics and Astronomy, Rutgers University, Piscataway New Jersey 08854, USA
[4] Massachusetts Institute of Technology, Department of Physics, Cambridge, MA, 02139, USA
[5] Department of Physics and Astronomy, Purdue University, West Lafayette, IN 47907, USA
[6] NYU-ECNU Institute of Physics at NYU Shanghai, 3663 Zhongshan Road North, Shanghai, 200062, China

† These authors contributed equally to this work
* To whom correspondence should be addressed; E-mail: lawray@nyu.edu.

# Abstract

The massless Dirac electrons found at topological insulator surfaces are thought to be influenced very little by weak, non-magnetic disorder. However, a resonance effect of strongly perturbing non-magnetic impurities has been theoretically predicted to change the dispersion and physical nature of low-energy quasiparticles, resulting in unique particle-like states that lack microscopic translational symmetry. Here we report the direct observation of impurities reshaping the surface Dirac cone of the model 3D topological insulator $Bi_2Se_3$. For the first time, a pronounced kink-like dispersion feature is observed in disorder-enriched samples, and found to be closely associated with the anomaly caused by impurity resonance in the surface state density of states, as observed by dichroic angle resolved photoemission spectroscopy (ARPES). The experimental observation of these features, which closely resemble theoretical predictions, has significant implications for the properties of topological Dirac cones in applied scenarios that commonly feature point defect disorder at surfaces or interfaces.

**Introduction**

Three dimensional topological insulators are bulk semiconductors with spin-helical Dirac cone surface states that span the bulk band gap [1,2]. Since their discovery around 2007 [3-6], the topological Dirac cone has appeared at the heart of a wide range of proposals for novel emergent quasiparticles and next-generation electronics, such as for the realization of exotic dyon-, axion-, or Majorana fermion-based physics [1,2,7-10]. Moreover, the spin-helical Dirac surface states are very robust against dilute non-magnetic impurities due to intrinsic immunities to backscattering and Anderson localization [11-16], and numerous studies have shown the surface Dirac cone remains qualitatively intact in the presence of weak non-magnetic disorder [17-19]. However, local probe studies have found that the effect of disorder on the *real space* electronic structure can be remarkably strong. Scanning tunneling microscopy (STM) experiments and complementary theoretical works have identified that even the simplest form of the non-magnetic impurity, a crystallographic point defect, will tend to give rise to resonance states very close to the Dirac point in 2D Dirac fermion systems such as topological insulators and graphene [20-27]. In systems like $Bi_2Se_3$, these surface defects are associated with distinctive patterns in STM topography images, and can redistribute local density of states (DOS) by building up a new peak tens of millielectronvolts above the surface Dirac point.

The relative significance of impurity resonances for real space electronic structure as opposed to momentum space is easy to understand if the impurities are low in density, because the associated resonance states will adhere locally to the sparse defects. However, if there is a sufficiently high density of non-magnetic impurities, the dispersion of the topological surface Dirac cone is expected to change significantly. Recent models have shown that when the density of impurities reaches an experimentally achievable high level, the impurity resonance states will behave collectively much like a flat band that hybridizes coherently with the upper Dirac cone, breaking the Dirac band into upper and lower branches [28, 29].

Measurements and theory suggest that electrons near the resonance adhere spatially to the disordered impurity lattice, which lacks translational symmetry. This scenario is at odds with the standard definition of a quasiparticle as a near-eigenstate of momentum. The momentum operator ($K$) is the generator of translations, and the spatial translation operator is defined as $T(\mathrm{x}) = \exp(-\mathrm{i}xK)$, implying that (near-) eigenstates of one operator should be (near-) eigenstates of the other. When an electron system is so disordered that translational symmetry in single-particle wavefunctions is absent on the length scale of one de Broglie wavelength, the result is termed a "bad metal", and it is assumed that the quasiparticle picture no longer applies. However, the resistance of topological surface electrons to Anderson localization and backscattering is thought to result in a unique quasiparticle-like character for electrons at the impurity resonance. Though these electrons profoundly lack translational symmetry, simulations suggest that their width cross-section in momentum space is narrower than one inverse wavelength [28, 29], and meets the nominal definition for a "good" quasiparticle. This is quite surprising, and means that inducing disorder at a topological insulator surface may enable the first experimental realization of an itinerant quasiparticle that propagates with a well-defined momentum, but occupies a spatial basis that lacks even near-neighbor translational symmetry. A quantitative correspondence with STM measurements of the spatially resolved DOS distribution has been cited as experimental evidence for this novel quasiparticle character [28]. However, the larger scale morphology of such highly disordered samples has been problematic for high resolution angle resolved photoemission (ARPES), and a topological Dirac cone reshaped by non-magnetic disorder has not been previously reported.

In this study, we present a high resolution linear-dichroic ARPES (LD-ARPES) investigation of defect-enriched $Bi_2Se_3$, and report the first experimental observation of a topological surface Dirac cone reshaped by non-magnetic impurity resonance.

The LD-ARPES spectra reveal an anomalous kink-like feature in the Dirac cone dispersion 40 meV above the Dirac point, which matches the predicted signature of coherent hybridization with an impurity resonance. Integrating over momentum shows that the dispersion anomaly is associated with a DOS peak, and that both features are progressively attenuated when successive intervals of low temperature annealing are applied to reduce disorder. All of these experimental results are highly consistent with theoretical predictions and with previous STM findings.

**Modeling impurity resonances in a topological surface Dirac cone**

Bismuth selenide ($Bi_2Se_3$) is widely seen as a model topological insulator system, with a single Dirac cone surface state connecting across a bulk band gap of roughly 0.3 eV. An idealized model of the ARPES spectral function of the linearly dispersive $Bi_2Se_3$ surface state is shown in Fig. **1a**. When a surface is simulated with randomly distributed non-magnetic impurities represented by scalar delta function potentials, the Dirac bands are broadened and their dispersion is changed (Fig. **1b**). The bands develop an apparent kink at E~40 meV above the Dirac point, corresponding with a blurred region in the simulated spectrum (dashed lines in Fig. **1b**). Weighting the spectra with the participation ratio (PR) of each eigenstate, a method to reveal spatially inhomogeneous states [30], highlights the impurity resonant states as they are more concentrated around the impurity sites (Fig. **1c**). In the weighted spectrum, the blurred feature then can be distinguished as being composed of two bands (white dashed lines) that are broken by the impurity resonance. The two disconnected dispersions are easier to distinguish at higher defect densities [28], and the corresponding features broaden as they approach the resonance energy of E~40 meV. Integrating the simulation over the 2D momentum space extracts the DOS, which shows a hump at the resonance energy (Fig. **1d**) where a local DOS peak has been observed for point-defects by STM [20-22].

**Results**:

Crystallographic point defects occur in typical bulk-grown $Bi_2Se_3$ samples with relatively low densities of ~<0.05% per 5-atom formula unit. In this study, defect-enriched $Bi_2Se_3$ bulk samples are synthesized by abbreviating the final annealing stage of sample growth (see Methods), following the procedure in Ref. [31]. X-ray diffraction (XRD) data (Fig. **2a**) and STM topography (Fig. **2b**) indicate that the resulting $Bi_2Se_3$ samples realize a large density of point-defects while maintaining good single-phase crystallinity. The resulting defect species are labeled on the STM topography map, based on previously identified correspondences [32]. The primary type of impurity is interstitial Se atoms residing on the outer surface or between the first and second $Bi_2Se_3$ quintuple-layers, with a combined density of $\rho$~0.08% per surface unit cell (i.e. per 0.15 $nm^2$ surface area). A small portion of anti-site $Bi_{Se}$ and Se vacancy defects are also observed with densities of $\rho$~0.03% and $\rho$~0.01%, respectively. This defect density has a good correspondence with the bulk Hall carrier density, but the precise numbers seen by STM fluctuate considerably from region to region (see Methods). The theoretical precondition for impurities significantly reconstructing the electronic structure is that there must be a defect population with local resonances at approximately the same energy $E_R$ relative to the Dirac point, and a density that exceeds a cutoff proportional to $(E_R)^2$ [28]. In the sparse-defect limit, $E_R$ is proportional to the negative inverse of the effective interaction strength between defects and the surface state ($U_{eff}$, defined in Methods) [27], suggesting that defects can be neglected if the ratio $\rho/(U_{eff})^2$ is significantly smaller than a critical threshold. Moreover, the surface state skin depth in $Bi_2Se_3$ and ARPES measurement depth are both limited to <~1 nm [33], meaning that the density of states contribution from deeper-lying resonances will not be strongly observed. Based on these considerations, the 2D density adopted for simulations is $\rho$=0.06%, representing the assumption that roughly half of the impurities observed in the top nanometer of the crystal will effectively share a common resonance energy, and collectively exceed the critical threshold.

High-resolution LD-ARPES was used to perform a targeted study of defect-derived changes in the surface electronic structure. Measurements were performed at the Advanced Light Source MERLIN beamline (BL 4.0.3), which also provides a small 50 µm beam profile to minimize feature broadening from macroscopic inhomogeneity. Two linearly polarized photoemission experimental geometries were selected as shown in the Fig. **2c-d**. The *σ*-polarization condition places the electric field parallel to the sample surface, and the π measurement condition polarization is inside the scattering plane, with a primarily out-of-plane (crystalline c-axis) orientation. Incident photon polarization was switched between π and *σ* geometries via control of the synchrotron beamline elliptically polarized undulator (EPU), to keep exactly the same beam spot on the sample. Previous ARPES measurements on the topological Dirac cone mainly used the π-pol experimental geometry, which gives strong emission from the primary $P_z$-orbital component of the Dirac cone. The *σ*-pol condition is rarely used due to its low efficiency. However, the weakness of emission from the electronic structure of a *pristine Dirac cone* is also a positive feature of this polarization condition, as the derivative term in the electron-photon interaction couples in-plane polarization directly to nanoscale in-plane inhomogeneity, which is greatest within defect resonance states and defect-rich regions of the sample surface. The asymmetric properties of these polarization matrix elements are reviewed in the Online Supplementary Material, and qualitatively associate the π and *σ* geometries with selective sensitivity to defect-poor and defect-rich surface regions, respectively.

High resolution ARPES measurements of the Dirac cone at a freshly cleaved sample surface are shown for the π-pol and *σ*-pol geometries in Fig. **3a**. The surface Dirac bands cross at the Dirac point and the ARPES matrix element effect gives the bands different intensity profiles under different polarizations. The lower Dirac cone is more difficult to trace due to the close proximity of the bulk valence band, and both the lower Dirac cone and the valence band show much higher intensity in *σ*-pol measurements. Momentum distribution curves (MDCs) of the Dirac bands are shown in Fig. **3e**, and show a large discrepancy between $\sigma$ and $\pi$ polarizations. At energies greater than E~>30 meV relative to the Dirac point, the $\sigma$-pol ARPES imaged Dirac bands always appear to be centered at smaller momenta, indicating a dispersion different from the states seen under $\pi$-pol. The MDCs of the simulated impurity-rich surface show a spectrum very similar to the Dirac cone imaged under $\sigma$-pol. In the simulation (Fig. **3d**), the Dirac bands have smaller momenta above the resonance energy (E~>30meV) within the kinked Dirac cone. This polarization dependence between the dispersion of ARPES-imaged Dirac bands is not possible for a perfectly homogenous sample, but the $\sigma$-pol dispersion closely resembles the emergent 'kink-like' feature associated with higher impurity densities. To reveal the quasiparticle dispersions more clearly, the MDCs of simulated and dichroic ARPES imaged Dirac cones ((Fig. **3d-e**) are fitted with Voigt functions to track their dispersions (Fig. **3h-i**). For simplicity**,** the 'kink-like' feature in the Dirac cone with resonance states was also treated as being composed of just two peaks for the purposes of the fitting procedure (See Supplementary Fig. 1 for the curve-by-curve fitting of MDCs). The $\sigma$- and $\pi$-pol dispersions have a relatively constant momentum offset at energies high above the Dirac point. At lower energies approaching the Dirac point, the $\sigma$-pol group velocity appears to become very large, as is common on the low-energy side of dispersion kinks, and the two dispersions merge rapidly beneath E~40-50meV. These anomalous features of the $\sigma$-pol dispersion can be closely reproduced by the simulation of a Dirac cone with dense impurities (Fig. **3h**).

Low temperature (LT) annealing provides a relatively safe way to mobilize the interstitial Se atoms in the van de Waals layer without creating new impurities of Se vacancy or $Bi_{Se}$ antisites. The temperature T~120°C is chosen to remain well below the thermal activation energy for creating new point defects, and beneath the T~>150°C low energy cutoff for eliminating larger morphological defects [34, 35]. To reduce disorder, the same sample was treated with intervals of LT annealing between synchrotron beamtimes, and re-cleaved for each new ARPES measurement. In a second experiment following one hour of LT annealing, ARPES data (Fig. **3b**) show significantly sharper bands, which is consistent with a reduced defect density. From the dichroic MDC comparison (Fig. **3f**), the discrepancy of dispersion between the π-pol and σ-pol imaged Dirac cone still exists, but is visibly smaller. The corresponding traced dispersion (Fig. **3j**) shows the same kink-like feature around E~40 meV in σ-pol imaged band structure, but with a smaller amplitude and an onset slightly closer to the Dirac point. The momentum offset between σ-pol and π-pol dispersions is reduced to roughly 60% of the value before annealing, and is less prominent outside of a narrow energy window from E~50-70 meV above the Dirac point. A final experiment followed 2 more hours of LT annealing (3 hours in total), and found almost no difference between the π-pol and σ-pol ARPES spectra (Fig. **3g**). Unlike the first 1hr LT anneal, very little change is noted in the sharpness of the bands following this final 2hr LT anneal. The traced bands (Fig. **3h**) show the same dispersion under both polarizations, and there is no such anomalous kink-like feature.

Though changes in band dispersion are of particular interest for physics and applications, a more basic property of impurity resonance known from STM and theory is the build-up of a DOS peak at the resonance energy [20-29]. This DOS feature cannot be identified from the ARPES band dispersions, as Luttinger's theorem does not apply to disordered systems [36-37]. To evaluate DOS, we instead sum the ARPES spectral function over the 2D momentum space, making use of the continuous rotational symmetry near the Dirac point to define $\text{DOS}(E) = \sum_k 2\pi|k| * I(E,k)$. This symmetrization procedure is applied separately to the σ-pol and π-pol dichroic ARPES data to obtain differently weighted approximations for the surface state DOS (Fig. **4a**). In all cases, the π-pol imaged DOS curve has a linear trend in the upper Dirac cone, matching expectations for a massless 2D Dirac fermion. However, the σ-pol data reveal an extra hump around E~40meV in the non-annealed base sample and after 1 hour of annealing. To exclude the possibility that this feature is from an extraneous matrix element effect, a corresponding simulation for a pristine Dirac cone is shown in Fig. **4b** based on the energy-resolved ARPES matrix elements identified in Ref. [38]. The symmetrized DOS curves for both measurement conditions show a good linear character, and neither has an anomalous peak feature, suggesting that the extra hump in the symmetrized DOS curves represents a real anomaly within the Dirac surface state. A cleaner view of the dichroic feature is obtained by subtracting the π-pol curve from the σ-pol DOS (Fig. **4c**), which reveals the anomalous hump as a peak at E~40meV in the dichroic DOS curve of the non-annealed sample. This peak continuously loses intensity with LT annealing, and is no longer identifiable after 3 hours of annealing. The PR-weighted impurity resonance peak simulated in Fig. **1d** is reproduced at the bottom of Fig. **4c**, and is energetically consistent with the dichroic DOS feature. This correspondence can only be qualitatively meaningful as it is filtered through ARPES matrix elements, however it indicates that the kinked band dispersion observed in σ-pol ARPES spectra is appropriately aligned with the impurity resonance in exactly the way that was theoretically predicted.

**Discussion**

These results show that $Bi_2Se_3$ samples with a high point defect density can exhibit significant changes in surface state dispersion and DOS, both of which are reversible via an aging process that includes low temperature annealing. Numerical modeling closely reproduces the experimental features, and suggests that the dispersion kink and higher energy momentum shift in σ-pol ARPES band structure originate from a small discontinuity in the Dirac bands due to hybridization with impurity resonance states. While this may seem like a disruption of the topological band connectivity, it does not involve time reversal symmetry breaking, and should not necessarily be viewed as a band gap. Band topology constraints are absent in the case of strong disorder, and the DOS distribution is peaked, rather than gapped, at the resonance energy.

With respect to earlier studies, it is noteworthy that the π-pol data *do not* show a similar kink-like dispersion feature, suggesting that the measurement conditions used for most ARPES experiments will give little weight to impurity-rich surface regions (see Supplementary Note 1). Moreover, the in-plane *σ*-pol orientation chosen for this study was selected to achieve particularly favorable matrix elements for resolving impurity physics (see Supplementary Note 2).

Unlike the conventional manifestation of a non-dispersive impurity band, the impact of disorder is seen to play out over a wide energy scale of several hundred millielectronvolts. Together with the theoretical protection against Anderson localization [28], this suggests that the Dirac cone electrons have a unique coherent relationship with the disordered impurity lattice, which preserves reasonably sharp quasiparticle-like character in spite of allowing large changes to the overall electronic structure. The DOS peak found at the dispersion kink is consistent with real-space STM investigations of impurity resonance, and has been predicted to greatly enhance the susceptibility to magnetic order in mean-field modeling for an ordered impurity lattice [27]. Further theoretical work will be needed to more fully understand the many-body susceptibilities of a dispersive quasiparticle-like mode that is profoundly inhomogeneous on a <100 nm scale, and for which the number of states in a `band' can not be treated as constant throughout momentum space (i.e. for which Luttinger's theorem does not hold). These results and predictions suggest that the non-magnetic impurities found ubiquitously at the surfaces and buried interfaces of real-world samples and devices may provide a far-reaching mechanism for shaping the physical properties of a TI surface.

**Methods**:

**1. High resolution linear dichroic ARPES measurements**

All ARPES measurements were performed at the BL4.0.3 MERLIN ARPES endstation at the Advanced Light Source, with a Scienta R8000 analyzer and base pressure better than $5\times10^{-11}$ Torr. The sample was maintained at T~20K, the time between s- and p-polarization measurements is roughly 4 hours, and surface band structure features were observed to be stable within ~10 meV on the time scale of the each LD-ARPES experiment (~<24 hours). The chemical potential for all samples was $E-E_D$~300 meV above the Dirac point, which matches the expectation of $E-E_D$=325 meV above the Dirac point for the Hall effect carrier density of $2.6\times10^{19}$ cm$^{-3}$. This surface potential is calculated using the band bending model in Ref. [39], which treats the surface state and bulk charge carriers on equal footing in the modified Thomas-Fermi Approximation (MTFA).

Surface doping of Bi$_2$Se$_3$ by adatoms and photon exposure is a natural concern in quantitative ARPES experiments, and multiple doping mechanisms can come into play [33,39-48]. However, the observed stability of the measured surface in this particular case is consistent with strong bulk screening associated with the high bulk carrier density, and with the observation that aging effects from residual gas and photon exposure tend to saturate at lower surface potentials of $E-E_D$~<290 meV [40]. The overall energy resolution was between 15-20meV, and photon flux was well below the regime on which photo-gating effects have been observed [41]. Measurements on the base sample were performed at hν=30 eV, and later post-annealing measurements were performed at hν=34 eV to obtain a higher photon flux, due to the rapid loss of photon throughput on the high energy grating beneath 50 eV.

## 2. Growth and STM characterization of defect-enriched Bi$_2$Se$_3$ samples:

High quality defect-enriched samples were grown following the procedure described in Ref. [31]. The resulting crystals show no sign of impurity phases. The Hall effect carrier density was 2.6×10$^{19}$ cm$^{-3}$, corresponding to an expected defect density of ~0.19 %, which corresponds reasonably well with the roughly ~0.12 % defect count in the top nanometer of the crystal seen by STM topography maps described in the main text. ARPES and STM measurements were performed close to the sample center, where the defect density is expected to be lower than on the periphery. The Hall carrier density and surface chemical potential are not greatly changed following the aging/annealing process A slight reduction in the surface chemical potential suggests that the bulk carrier density is reduced by 10-20% in later measurements, relative to density beneath the beam spot in the initial measurement. Additionally, it is expected that the post-annealed defect distribution may be more homogeneous across microcrystalline domains.

STM data were obtained at low temperature (T<50K) with tips calibrated on an Au(111) surface. The topographic map in Fig. **2b** was measured using a bias of -0.7 V and a tunneling current of 100 pA. Rapid topographical maps of multiple ~100 nm regions were sampled, and were consistent with previous analyses that suggest a stochastically random placement of impurities within local regions (see the supplemental material of Ref. [28]). However, the standard deviation in defect density for different surface regions separated by several microns was much higher than the variation expected from the Poisson distribution. This variation reveals large fractional differences (up to a factor of ~2) in the local impurity densities that are averaged over in the ARPES spectral function.

**3. Modeling the TI surface**

The surface state is modeled as a spin-helical 2D Dirac cone on a hexagonal lattice, perturbed by scalar delta-function-like impurities, as described in Ref. [28]. The impurity potential is effectively reduced by the fact that the delta-function-like potential is not well resolved in the model, which imposes a high energy cutoff on the kinetic basis for state diagonalization. An effective value for the potential is calculated as the trace of the potential Hamiltonian for a single impurity ($U_{eff}$=Tr($H_U$)=-1 eV), whereas the uncorrected potential would have a value of U=-35 eV, as defined in Ref. [28]. The impurities are randomly distributed with a density of $\rho$ =0.06%. Spectral features were convoluted by a Lorentzian function with 30 meV peak width at half maximum, except where otherwise noted.

Intensities in Fig. **1c** are weighted by the participation ratio *P* of each single-particle eigenstate. The participation ratio gives higher intensity for states that are less evenly distributed throughout space, and is used to highlight emission from defect resonance states. It is defined as:

$$P_\alpha = \sum_i n_{\alpha,i}^2$$

where the sum is over all sites in the system, and $n_{\alpha,i}$ is the local DOS on site *i* of |α⟩, which is an eigenstate of the full Hamiltonian.

References:

[1] Hasan, M. Z. & Kane, C. L. Colloquium: Topological insulators. Rev. Mod. Phys. **82**, 3045 (2010).
[2] Qi, X-L. & Zhang, S-C. Topological insulators and superconductors. Rev. Mod. Phys. **83**, 1057 (2011)
[3] Fu, L., Kane, C. L. & Mele, E. J. Topological insulators in three dimensions. Phys. Rev. Lett. **98**, 106803 (2007).
[4] Zhang, H., Liu, C.-X., Qi, X-L., Dai, X., Fang, Z. & Zhang, S-C. Topological insulators in $Bi_2Se_3$, $Bi_2Te_3$ and $Sb_2Te_3$ with a single Dirac cone on the surface. Nat. Phys. **5**, 438-442 (2009).
[5] Xia, Y. et al. Observation of a large-gap topological-insulator class with a single Dirac cone on the surface. Nat. Phys. **5**, 398-402 (2009).
[6] Chen, Y. L. et al. Experimental realization of a three-dimensional topological insulator, $Bi_2Te_3$. Science **325**, 178-181 (2009)
[7] Qi, X-L., Li, R., Zang, J. & Zhang, S-C. Inducing a magnetic monopole with topological surface states. Science **323**, 1184-1187 (2009)
[8] Yu, R. et al. Quantized anomalous Hall effect in magnetic topological insulators. Science **329**, 61 (2010).
[9] Chang, C.-Z. et al. Experimental observation of the quantum anomalous Hall effect in a magnetic topological insulator. Science **340**, 167 (2013).
[10] Fu, L. & Kane, C. L. Superconducting proximity effect and Majorana fermions at the surface of a topological insulator. Phys. Rev. Lett. **100**, 096407 (2008)
[11] Hsieh, D. et al. Observation of Unconventional quantum spin textures in topological linsulators. Science **323**, 919-922 (2009).
[12] Fu, L. & Kane, C. L., Topology, delocalization via average symmetry and the symplectic Anderson transition. Phys. Rev. Lett. **109**, 246605 (2012).
[13] Roushan, P. et al. Topological surface states protected from backscattering by chiral spin texture. Nature **460**, 1106-1109 (2009).
[14] Zhang, T. et al. Experimental Demonstration of Topological surface states protected by time-reversal symmetry. Phys. Rev. Lett. **103**, 266803 (2009).
[15] Alpichshev, Z. et al., STM Imaging of electronic waves on the surface of $Bi_2Te_3$: topologically protected surface states and hexagonal warping effects. Phys. Rev. Lett. **104**, 016401 (2010).
[16] Beidenkopf, H. et al., Spatial fluctuations of helical Dirac fermions on the surface of topological insulators. Nature Physics **7**, 939–943 (2011)
[17] Hsieh, D. et al. A tunable topological insulator in the spin helical Dirac transport regime. Nature **460**, 1101–1105 (2009).
[18] Wray, L. A. et al. Observation of topological order in a superconducting doped topological insulator. Nat. Phys. **6**, 855859 (2010).

[19] Valla, T., Pan, Z.-H., Gardner, D., Lee, Y. S., Chu, S. Photoemission spectroscopy of magnetic and nonmagnetic impurities on the surface of the $Bi_2Se_3$ topological insulator. Phys. Rev. Lett. **108**, 117601 (2012).

[20] Alpichshev, Z. et al. STM imaging of impurity resonances on $Bi_2Se_3$. Phys. Rev. Lett. **108**, 206402 (2012).

[21] Teague, M. L. et al. Observation of Fermi-energy dependent unitary impurity resonances in a strong topological insulator $Bi_2Se_3$ with scanning tunneling spectroscopy. Solid State Comm. **152**, 747-751 (2012).

[22] Ugeda, M. M., Brihuega, I., Guinea, F. & GomezRodriguez, J. M. Missing atom as a source of carbon magnetism. Phys. Rev. Lett. **104**, 096804 (2010).

[23] Wehling, T. O. et al. Local electronic signatures of impurity states in grapheme. Phys. Rev. B **75**, 125425 (2007)

[24] Biswas, R. R. & Balatsky, A. V. Impurity-induced states on the surface of three-dimensional topological insulators. Phys. Rev. B **81**, 233405 (2010).

[25] Black-Schaffer, A. M. & Balatsky, A. V. Subsurface impurities and vacancies in a three-dimensional topological insulator. Phys. Rev. B **86**, 115433 (2012).

[26] Black-Schaffer, A. M. & Balatsky, A. V., Strong potential impurities on the surface of a topological insulator. Phys. Rev. B **85**, 121103(R) (2012).

[27] Black-Schaffer, A. M. & Yudin, D., Spontaneous gap generation on the surface of weakly interacting topological insulators using nonmagnetic impurities. Phys. Rev. B **90**, 161413(R) (2014).

[28] Xu, Y. et al. Disorder enabled band structure engineering of a topological insulator surface. Nat. Comm. **8**, 14801 (2017).

[29] Zhong, Min. et al. Effect of impurity resonant states on optical and thermoelectric properties on the surface of a topological insulator. Sci. Rep. **7**, 3971 (2017)

[30] Thouless, D. J., Electrons in disordered systems and the theory of localization. Phys. Rep. **13C**, 93-142 (1974).

[31] Petrushevsky, M. et al. Probing the surface states in $Bi_2Se_3$ using the Shubnikov-de Haas effect. Phys. Rev. B **86**, 045131(R) (2012).

[32] Dai, J. et al. Toward the intrinsic limit of the topological insulator $Bi_2Se_3$. Phys. Rev. Lett **117**, 106401 (2016).

[33] Wray. L. A. et al. Spin-orbital ground states of superconducting doped topological insulators: A Majorana platform. Phys. Rev. B **83**, 224516 (2011)

[34] Xue, L. et al. First-principles study of native point defects in $Bi_2Se_3$. AIP Advances **3**, 052105 (2013).

[35] Li, H. D. et al. The van der Waals epitaxy of $Bi_2Se_3$ on the vicinal Si(111) surface: an approach for preparing high-quality thin films of a topological insulator. New J. Phys. **12** 103038 (2010).

[36] Luttinger, J. M. & Ward, J. C. Ground-state energy of a many-fermion system. II. Physical Review **118**, 1417–1427 (1960).

[37] Luttinger, J. M. Fermi surface and some simple equilibrium properties of a system of interacting fermions. Physical Review **119**, 1153–1163 (1960).

[38] Cao, Y. et al. Mapping the orbital wavefunction of the surface states in three-dimensional topological insulators. Nat. Phys. **9**, 499-504 (2013).
[39] Wray, L. A. et al. Chemically gated electronic structure of a superconducting doped topological insulator system, J. Phys.: Conf. Series 449, 012037 (2013).
[40] Frantzeskakis, E. et al. Trigger of the ubiquitous surface band bending in 3D topological insulators. Phys. Rev. X **7**, 041041 (2017)
[41] Kordyuk, A. A. et al. Photoemission-induced gating of topological insulators. Phys. Rev. B **83**, 081303 (2011).
[42] Wray, L. A. et al. A topological insulator surface under strong Coulomb, magnetic and disorder perturbations, Nature Physics **7**, 32–37 (2011)
[43] Bianchi, M. et al., Coexistence of the topological state and a two-dimensional electron gas on the surface of $Bi_2Se_3$. Nat. Commun. **1**, 128 (2010).
[44] Benia, H. M., Lin, C., Kern, K. and Ast, C. R. Reactive chemical doping of the $Bi_2Se_3$ topological insulator. Phys. Rev. Lett. **107**, 177602 (2011).
[45] Bahramy, M. S. et al. Emergent quantum confinement at topological insulator surfaces Nat. Commun. **3**, 1159 (2012).
[46] Valla, T., Pan, Z.-H., Gardner, D., Lee, Y. S. & Chu, S. Photoemission spectroscopy of magnetic and nonmagnetic impurities on the surface of the $Bi_2Se_3$ topological Insulator Phys. Rev. Lett. **108**, 117601 (2012).
[47] Chen, C. et al. Robustness of topological order and formation of quantum well states in topological insulators exposed to ambient environment. Proc. Natl. Acad. Sci. USA. **109**, 3694 (2012).
[48] Jiang, R. et al. Reversible tuning of the surface state in a pseudobinary Bi 2 (Te-Se) 3 topological insulator. Phys. Rev. B **86**, 085112 (2012).

Acknowledgements:

We are grateful for discussions with P. Chaikin, Y.-D. Chuang, D. Huse, and P. Moon. This research used resources of the Advanced Light Source, which is a DOE Office of Science User Facility under contract no. DE-AC02-05CH11231. R.R.B. was supported by Purdue University startup funds. Synthesis and analysis instrumentation at NYU is supported by NSF under MRI-1531664, and from the Gordon and Betty Moore Foundation's EPiQS Initiative through Grant GBMF4838. Work at NYU was supported by the MRSEC Program of the National Science Foundation under Award Number DMR-1420073. The STM work at Rutgers is supported by NSF under grant DMR-1506618.

Authors Contribution:

L.M. and Y.X. carried out the ARPES experiments with support from D.O., S.A.B., E.K., H.H., and J.D.; STM measurements were performed by W.Z., with guidance from W.W.; high quality defect-enriched samples were developed by T.S. and J.C.; simulations were performed by Y.X. with guidance from R.R.B.; L.M., Y.X., W.Z., and L.A.W participated in the analysis, figure planning and draft preparation; L.A.W. was responsible for the conception and the overall direction, planning and integration among different research units.

Figure 1

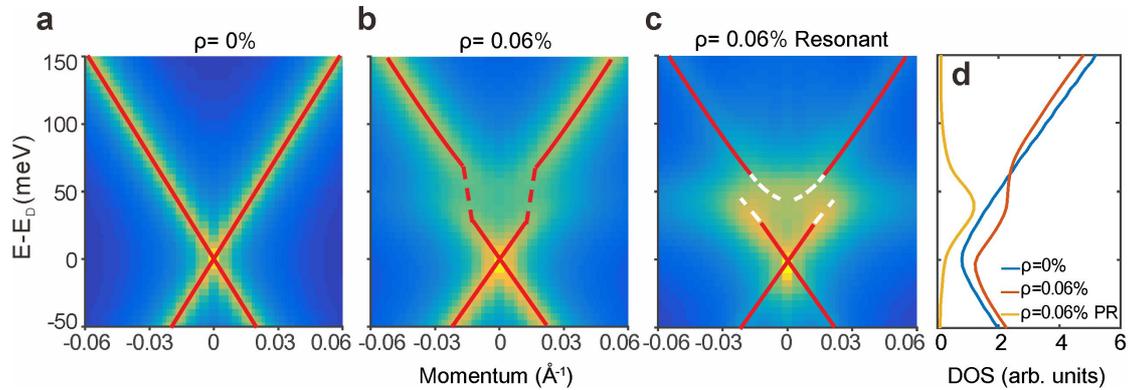

**Figure 1**. **Impurity resonance in momentum space. a**, A simulated $Bi_2Se_3$ surface Dirac cone without impurities. Red lines trace the linear band dispersion. **b**, A simulated Dirac cone reshaped by scalar impurities ($U_{eff}$ =1 eV, density ρ=0.06% per 2D surface unit cell). A kink-like dispersive feature occurs at the impurity resonance energy, and is traced with dashed lines. **c**, The simulated Dirac cone is weighted by participation ratio to highlight impurity-resonant states. The kink-like feature is resolved more clearly as containing the split dispersions predicted in Ref. [28] (white dashed lines). **d**, The corresponding simulated DOS curves of (blue) a pristine Dirac cone, (red) a Dirac cone with impurities, and (yellow) a participation ratio (PR) weighted simulation with impurities.

Figure 2

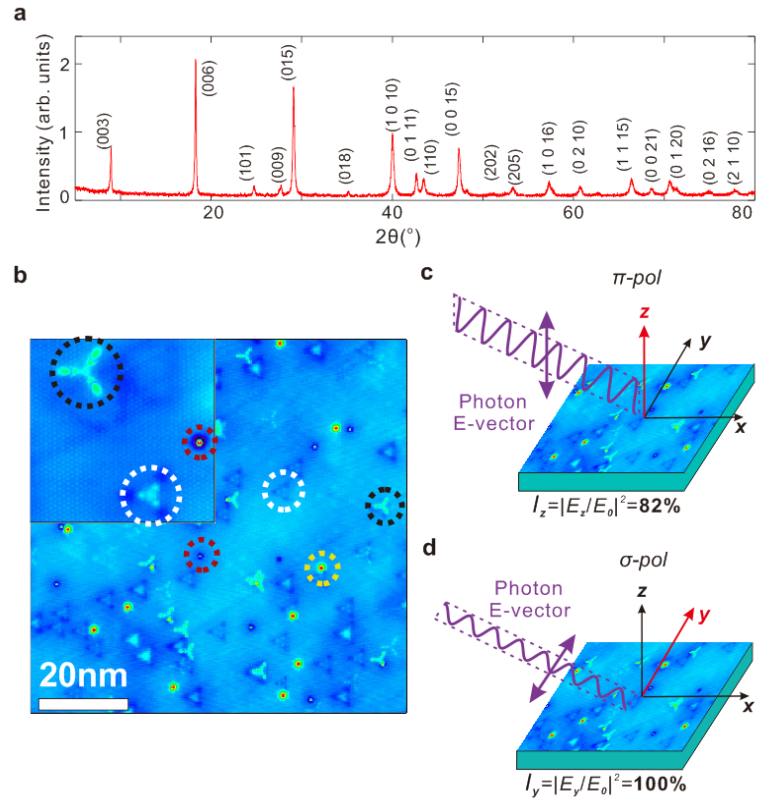

**Figure 2**. **Defect-enriched Bi$_2$Se$_3$. a**, XRD from the defect-enriched Bi$_2$Se$_3$ sample, showing no impurity phase features. **b**, STM topography of an 80×80 nm$^2$ cleaved Bi$_2$Se$_3$ surface with lattice defects/impurities. The defects are predominantly excess Se (red circles) on the top surface and (white circles) between first and second quintuple layers. (yellow and black circles) Antisite Bi$_{Se}$ defects are also common. An expanded inset shows distinctive defect profiles used for characterization. **c**, The π-polarization ARPES measurement geometry, with the electric field of incident photons mostly normal to the sample surface (projecting 82% onto the z-axis). **d**, The σ- polarization ARPES measurement geometry gives an electric field that projects 100% onto the in-plane y-axis (the ARPES analyzer slit axis).

Figure 3

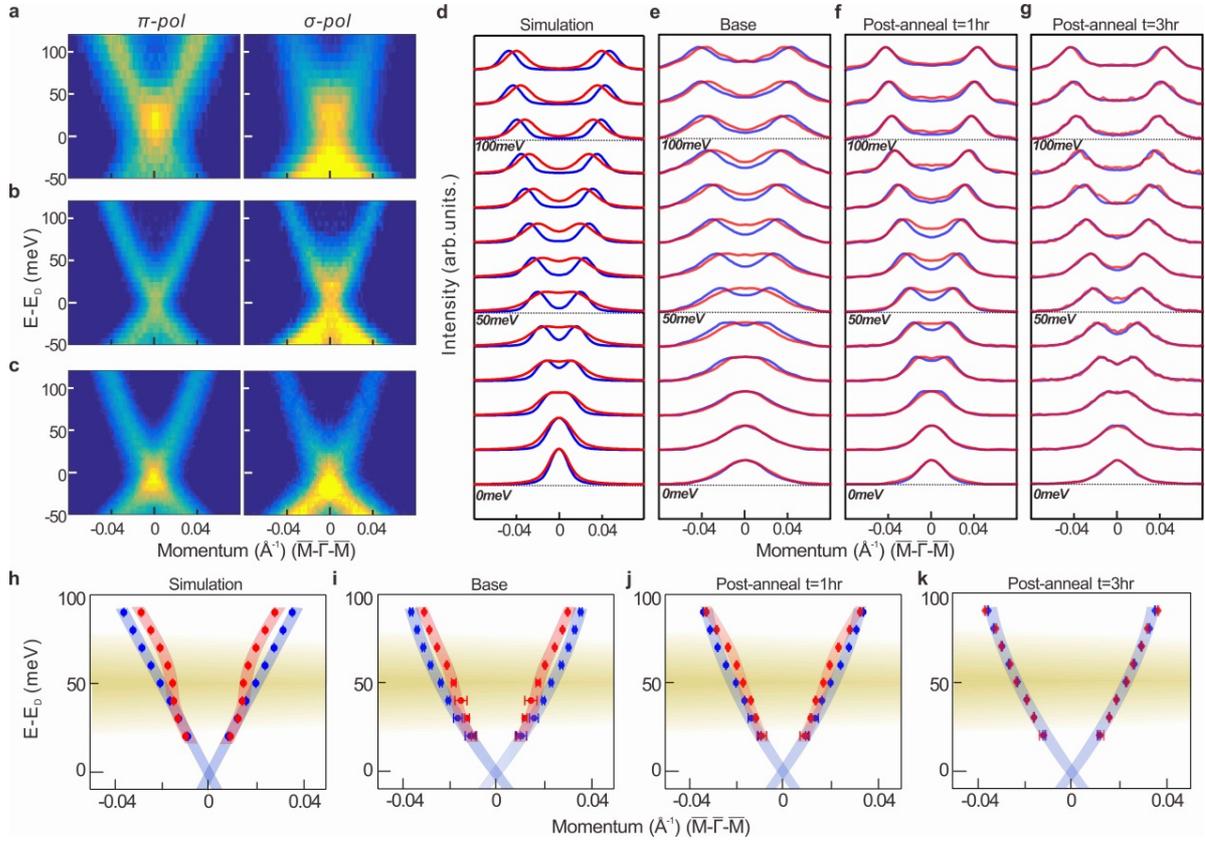

**Figure 3. Measuring the defect-derived dispersion anomaly.** Symmetrized π-pol and σ-pol ARPES images of Dirac cone band structure are shown for the same defect-enriched sample (**a**) before LT annealing and after (**b**) 1 hour and (**c**) 3 hours of LT annealing. **c**, The MDCs of a simulated Dirac cone (red) with and (blue) without impurities are extracted from Fig. 1**a-b**, and shown with a 10 meV step, starting from the Dirac point (set to 0 energy). To facilitate comparison, all MDCs are normalized to the same amplitude. **e-f**, The corresponding MDCs of panels (**a-c**) are shown with (red) σ-pol and (blue) π-pol. **h-k**, Dispersions are obtained from fitting the red and blue coded MDCs in panels **d-g** with two Voigt functions. A horizontal shaded region indicates the expected defect resonance energy [20,28].

Figure 4

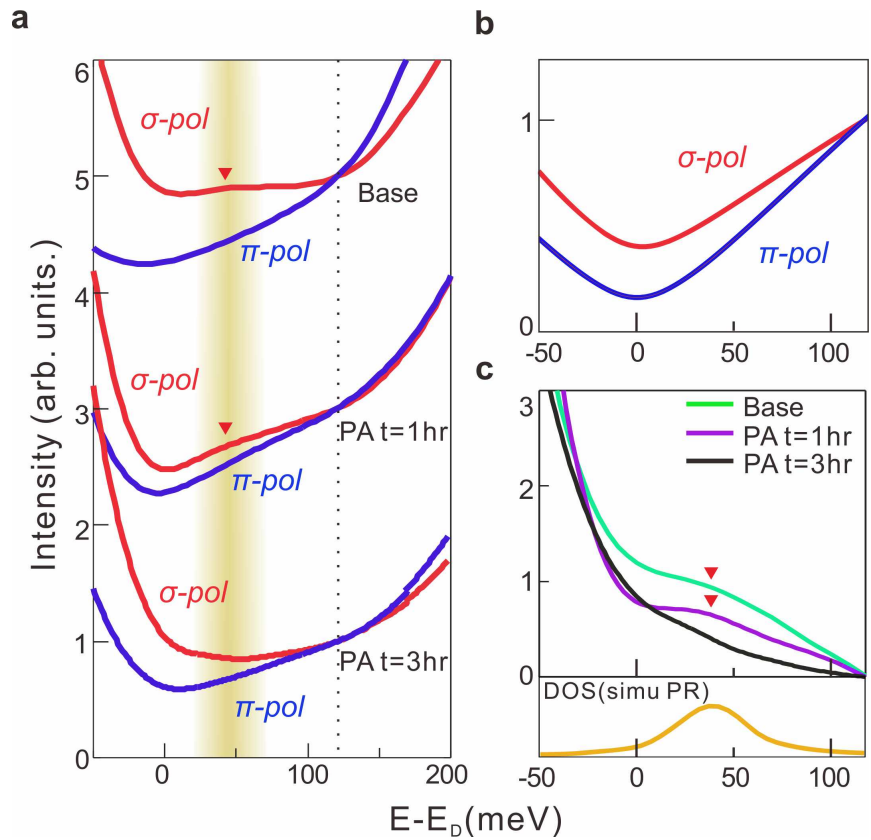

**Figure 4. The impurity resonance DOS peak. a**, ARPES DOS curves estimated from (red) σ-pol and (blue) π-pol are shown for different levels of LT post-annealing (PA). The curves are offset in increments of 2, and normalized at E=120 meV above the Dirac point (dashed line), an energy that is higher than (shaded region) the expected impurity resonance and low enough to avoid the bulk conduction band. **b**, Simulated ARPES DOS curves for a pristine Dirac cone, based on empirical photoemission matrix elements from a Ref. [38]. **c**, Dichroic DOS curves the subtracting the π-pol ARPES DOS from σ-pol ARPES DOS. The dichroic DOS curves are compared with (bottom) the PR-weighted impurity resonance simulation from Fig. 1d. The anomalous peak-like feature that vanishes with annealing is indicated with filled triangles.

# Supplementary Material:

# Observation of a Topological Insulator Dirac Cone Reshaped by Non-magnetic Impurity Resonance

Lin Miao,[1,2][†] Yishuai Xu,[1][†] Wenhan Zhang,[3][†] Daniel Older,[1]

S. Alexander Breitweiser,[1] Erica Kotta[1], Haowei He,[1] Takehito Suzuki,[4]

Jonathan D. Denlinger,[2] Rudro R. Biswas,[5]

Joseph Checkelsky,[4] Weida Wu,[3] L. Andrew Wray,[1,6]∗

[1] Department of Physics, New York University, New York, New York 10003, USA
[2] Advanced Light Source, Lawrence Berkeley National Laboratory, Berkeley, CA 94720, USA
[3] Rutgers Department of Physics and Astronomy, Rutgers University, Piscataway New Jersey 08854, USA
[4] Massachusetts Institute of Technology, Department of Physics, Cambridge, MA, 02139, USA
[5] Department of Physics and Astronomy, Purdue University, West Lafayette, IN 47907, USA
[6] NYU-ECNU Institute of Physics at NYU Shanghai, 3663 Zhongshan Road North, Shanghai, 200062, China

† These authors contributed equally to this work
∗ To whom correspondence should be addressed; E-mail: lawray@nyu.edu.

**Supplementary Note 1**: Photon polarization as a filter for impurity-resosnt or non-resonant states

In this investigation, σ-polarization (in-plane) is used to obtain a higher sensitivity to electrons that have a spatial distribution strongly influenced by defects, and π polarization (mostly surface-normal) is used to measure electronic states that are less defect-resonant and better resemble an ideal topological Dirac cone. This association is justified from theoretical and empirical considerations outlined below. Adopting the dipole approximation and single-step photoemission picture (standard for the photon energies used in this study), the photoemission intensity is determined by the following matrix element [1]:

$$I(f) \sim |\langle f | \nabla \cdot \mathbf{A} | \psi \rangle|^2 \qquad (S1)$$

where **A** is the vector potential of the incident beam, and $|\psi\rangle$ is an occupied single-electron state inside the material. The final state $|f\rangle$ is assumed to have the spatial form of a free particle state, and to overlap with the topmost ~1nm of the material. Moreover, this state is assumed to be orthogonal to $|\psi\rangle$, so that elements of the photon perturbation that commute with the unperturbed Hamiltonian of the system can be neglected.

**With σ-polarization, photoemission is sensitive to in-plane structure:** For an experiment performed with σ-polarization on a fully homogeneous material with a 2-dimensional surface, the photon perturbation commutes with the kinetic part of the Schrödinger equation, and thus has no cross section for photoemission. Introducing in-plane inhomogeneity in the form of random point defects will create a photoemission signal from spatially structured electronic states that have a significant probability of being found in the inhomogeneous region (such as defect resonance states).

The two implications of this idealized picture are that the signal from resonance states may be enhanced with σ-polarization, and that the signal from non-resonance states should be weak, giving improved signal to noise for observing defect resonance. The second point is not generically true when one considers real materials, but turns out to be valid for our measurements, as noted in the main text. In fact, data acquisition times for σ polarization were approximately 5-10 times as long to achieve comparable statistics to π-polarization. The symmetries introduced via defect-mediated hybridization with bulk band structure are expected to further increase sensitivity to defects under σ-polarization, as the closest band in energy is the valence band, which has very strong intensity under σ-polarization (visible in the intensity tail from large binding energy in Fig. 4a of the main text).

On an inhomogeneous surface, filtering for defect-resonant states will increase the fractional spectral weight from high-defect-density regions. In this context, it is noteworthy that the density of impurities seen by STM varied by up to a factor of 2 in different regions studied within the area of an ARPES beam spot (see Methods 2), and this inhomogeneity may cause much of the surface to be beneath the critical impurity density threshold for achieving the reconstruction of spectral features. The weight from surface regions with impurity densities above the threshold is expected to be higher under σ- polarization.

**With π-polarization, photoemission near the Brillouin zone center is suppressed by in-plane point defects**: For the case of π-polarization, it is important to note that the ARPES matrix element includes a projection onto the basis of homogeneous free particle states $|f\rangle$. Non-defect-resonant states of the topological Dirac cone are composed of long-wavelength wavefunction components that are almost entirely found in the momentum space region mapped by ARPES in this investigation. However, resonance states have a significant admixture of short-wavelength wavefunction components to which our ARPES scans of the k<~0.1 Å$^{-1}$ Dirac cone momentum region are blind (this can be seen from the sharply contoured real-space structures seen around defects by STM). This means that impurity-resonant electrons can be expected to count less towards π-pol ARPES maps of the Dirac cone, even if a defect-resonant state has a k-space intensity maximum that overlaps with the Dirac cone.

**Supplementary Note 2**: **Simulating the effect of ARPES polarization matrix elements on the DOS measurement for a defect-free Dirac cone:**

The Fig. 4(b) in the main text presents simulated ARPES DOS curves of the $Bi_2Se_3$ Dirac cones, based on matrix elements extracted from previous systematic ARPES measurements [2]. For this simulation, the gapless dispersion near the Dirac point is set by a Dirac velocity of $v_D$=2.6 eV-Å, and the ARPES intensity ($I_{avg}(E)$) under $\sigma$-polarization is obtained directly from the Ref. [2] anisotropy curves for energies of -25, 0, 50, 100, and 200 meV, and interpolated for intermediate energy values. The $\pi$-polarization intensity trend is taken to be flat as a function of energy, consistent with the cleaner linear trend seen in Dirac cone DOS estimates based on this measurement geometry. Gauss and Lorentz broadening are set to 0.0162 $Å^{-1}$ peak width at half maximum in momentum space, and on the energy axis are set to 20 and 40 meV, respectively. Off-axis intensity in momentum space was described by the cos-like trends identified in Ref. [2], and contributes to the simulated spectral intensity due to the isotropic momentum broadening parameters.

With respect to the factors discussed in Supplementary Note 1, the $\sigma$-polarization measurement geometry corresponds to an azimuthal angle of 90$^o$ in Ref. [2], and minimizes intensity from the upper Dirac cone as desired for optimal sensitivity to impurity resonance. This polarization condition also provides a strong photoemission matrix element for the energetically close valence band, which is expected to contribute to the partial DOS of impurity resonance states.

**Supplementary Note 3**: **The symmetrization of LD-ARPES images and related MDCs.**

$Bi_2Se_3$(111) has three-fold symmetry ($C_3$) about the surface-normal axis. The ARPES measurement cuts in our research were along the 2D M-Γ-M direction, which means there is strong left-right matrix element asymmetry on the momentum axis of photoemission spectra (Supplementary Fig. 2). The interplay of these matrix elements with the continuum nature of the impurity resonance states will lead to a slight left-right asymmetry in the apparent dispersion of the surface state on either side of the Dirac point, as can be seen in Supplementary Fig. 2.

While this asymmetry provides support for the attribution of impurity physics, it carries the downside of adding degrees of freedom when fitting band structure, and giving a very asymmetrical intensity profile that increases the dynamic color range needed for image plots. To achieve a clearer dichroic comparison, a left-right symmetrization has been applied to the 2D images and MDCs shown in the main text. This does not impact the s- vs. p- polarization comparison, as the measurement conditions (including beam spot) are identical. The analysis of Fig. 3-4 in the main text is robust even when a deliberate centering error is introduced in the symmetrization process.

**Supplementary Note 4**: **Sample annealing and cleaving:**

For each measurement, the sample was top-posted in the atmosphere, and cleaved and measured *in situ* at low temperature (~20K) under ultra-high vacuum (<$5*10^{-11}$ torr). Annealing was performed in atmosphere between synchrotron-based experiments, which were separated by 4 month intervals.

Following each annealing and cleaving process, the Fermi level was found to be approximately 300meV above the Dirac point with a variation within 20meV. Estimated Fermi energies vary from an initial value of 305 meV to a minimum of 285 meV (see Supplementary Fig. 2), suggesting a possible loss of up to 20% of bulk carriers in post-annealed measurements. The fitting error in determination of the

Dirac point energy is estimated by less than 10 meV, from one cleavage to the next. This error does not factor into the linear dichroic differences within a given cleavage.

References

[1] Damascelli, A., Hussain, A. & Shen, Z.-X., Angle-resolved photoemission studies of the cuprate superconductors, Rev. Mod. Phys. 75, 473 (2003).

[2] Cao, Y. et al. Mapping the orbital wavefunction of the surface states in three-dimensional topological insulators, Nat. Phys. 9, 499-504 (2013).

**Supplementary Figure 1**:

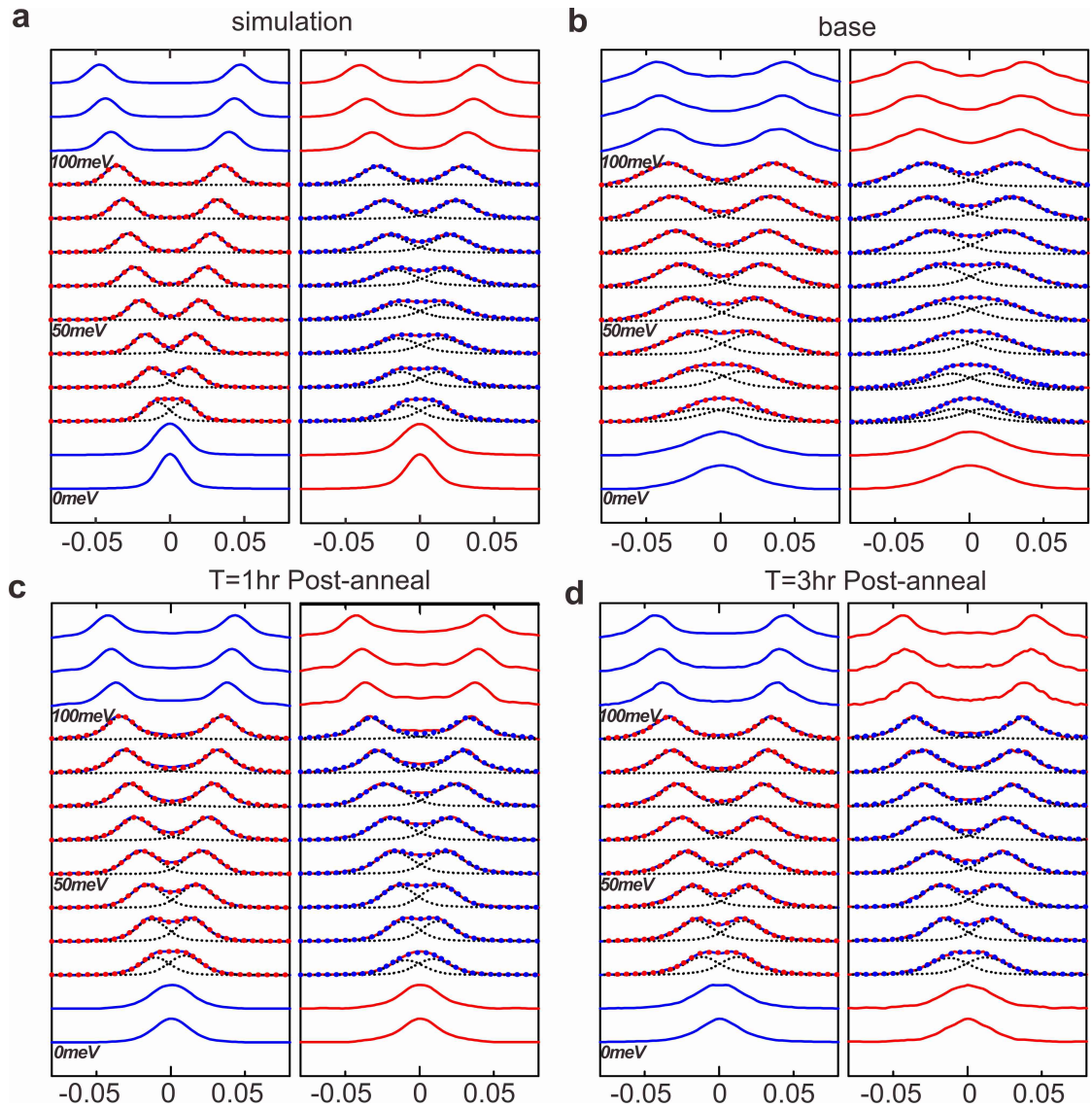

**Figure S1. Curve-by-curve fitting of Fig. 3 dispersions**. **a,** MDCs of the simulated Dirac cone spectra from Fig. 3 of the main text (blue) without and (red) with defects are fitted with two Voigt profiles (black dashed) to track the effective Dirac band dispersions. The same fitting procedure is applied to the MDCs of dichroic ARPES spectra **b,** without LT annealing, **c,** with 1hr LT post-annealing and **d,** with 3 hr LT post-annealing. In (**b**-**d**), the blue curves are from π-pol measurements and red curves are from σ-pol measurements. Only the MDCs with energy E>=20 meV above Dirac point provide sufficiently resolved features for accurate fitting.

**Supplementary Figure 2**:

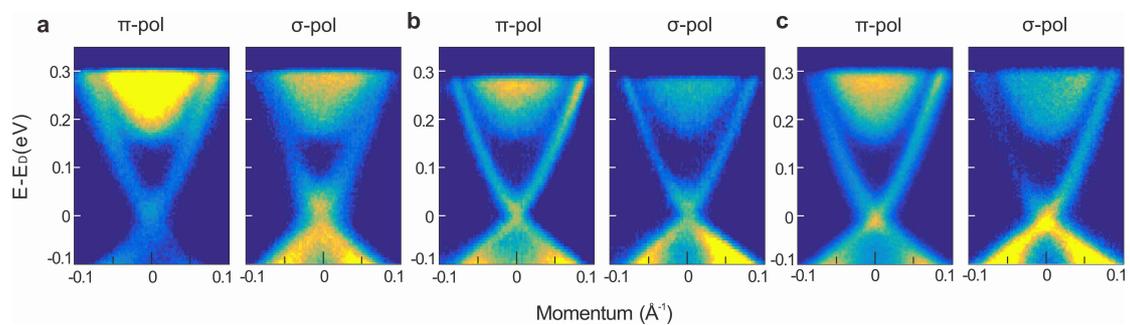

**Figure. S2. The raw data of LD-ARPES measurements**. Raw LD-ARPES data showing the full upper Dirac cone (a) before annealing, (b) after 1 hour of annealing, and (c) after 3 hours of annealing.